\documentclass{article}

\usepackage{arxiv}

\usepackage[utf8]{inputenc} 
\usepackage[T1]{fontenc}    
\usepackage{hyperref}       
\usepackage{url}            
\usepackage{booktabs}       
\usepackage{amsfonts}       
\usepackage{nicefrac}       
\usepackage{microtype}      
\usepackage{graphicx}
\usepackage{cite}
\usepackage{doi}

\usepackage{bm}
\usepackage{amsmath}
\usepackage{algorithm}
\usepackage{algpseudocode}
\DeclareMathOperator*{\argmin}{arg\,min}
\usepackage[table]{xcolor}
\usepackage{changepage}
\usepackage{multirow}

\usepackage{listings}
\title{A Multi-step Approach for Minimizing Risk in Decentralized Exchanges\\
\vspace{0.1cm}
\large{The SIAG/FME Code Quest 2023 Winning Strategy}}

\author{Daniele Maria Di Nosse\thanks{These authors contributed equally to this work.} \\
	Scuola Normale Superiore\\
	Pisa, Italy \\
	\texttt{daniele.dinosse@sns.it} \\
	\And
	Federico Gatta\footnotemark[1] \\
	Scuola Normale Superiore\\
	Pisa, Italy \\
	\texttt{federico.gatta@sns.it} \\
}

\renewcommand{\shorttitle}{\textit{arXiv} Template}

\hypersetup{
pdftitle={A Multistep Approach for Minimizing Risk in Decentralized Exchanges},
pdfsubject={q-bio.NC, q-bio.QM},
pdfauthor={Daniele Maria~Di Nosse, Federico~Gatta},
pdfkeywords={Decentralized Exchanges, Conditional Value at Risk},
}

\date{}

\begin{document}
\maketitle

\begin{abstract}
    Decentralized Exchanges are becoming even more predominant in today's finance. Driven by the need to study this phenomenon from an academic perspective, the SIAG/FME Code Quest 2023 was announced. Specifically, participating teams were asked to implement, in Python, the basic functions of an Automated Market Maker and a liquidity provision strategy in an Automated Market Maker to minimize the Conditional Value at Risk, a critical measure of investment risk. As the competition's winning team, we highlight our approach in this work. In particular, as the dependence of the final return on the initial wealth distribution is highly non-linear, we cannot use standard ad-hoc approaches. Additionally, classical minimization techniques would require a significant computational load due to the cost of the target function. For these reasons, we propose a three-step approach. In the first step, the target function is approximated by a Kernel Ridge Regression. Then, the approximating function is minimized. In the final step, the previously discovered minimum is utilized as the starting point for directly optimizing the desired target function. By using this procedure, we can both reduce the computational complexity and increase the accuracy of the solution. Finally, the overall computational load is further reduced thanks to an algorithmic trick concerning the returns simulation and the usage of Cython.
\end{abstract}


\section{Introduction}
Nowadays, Decentralized Finance and Crypto markets are hot topics in the financial world. Every day, billions of dollars are traded in Decentralized Exchanges (DEXs) \cite{defillama}. This astonishing growth is related to the intrinsic weakness of Centralized Exchanges, mainly associated with the faith in the centralized party \cite{schar2021decentralized}, which has sometimes led to financial disasters, as in the case of Futures Exchange (FTX). To overcome the need for an intermediary, DEXs exploit special programs called smart contracts. A smart contract is a set of immutable instructions that dictate how crypto assets are handled on the exchange and how the price is formed \cite{alotaibi2021smart}. Specifically, the functioning of the more recent DEXs is dictated by the so-called Automated Market Makers (AMMs).\\
\linebreak
AMMs are algorithms that handle the exchange's liquidity and determine the asset price. In more detail, they act via liquidity pools and rules for price formation. In the simplest case, liquidity pools are money boxes with two coins. Liquidity Providers (LPs) and Liquidity Takers (LTs) are the main actors. The former provides liquidity by adding both coins to the pool. The latter takes the liquidity by swapping the coins. The LPs' contribution to the pool is witnessed by the Liquidity Providing coins (LP coins), representing the fraction of the pool owned by each LP. On the other hand, LTs pay the service with a fee, which is then distributed to all the LPs according to the number of LP coins they possess and the protocol itself. The distribution rule of this fee is protocol-specific. As for the price rules, they are a set of mathematical functions that regulate the swapping process. In particular, the most relevant approach is the Constant Product Market Maker (CPMM), where the price of a coin is determined to keep the product of the coin reserves, before and after the swap, constant. Finally, it is noteworthy to point out that, often, there are more liquidity pools for the same couple of tokens that act independently, giving rise to an optimization problem for the LP agents longing to maximize their profit. Due to the strong interest in the design of optimal investment strategies for LPs operating in multiple pools (see, for example, \cite{cartea2023decentralised}), the SIAM Activity Group on Financial Mathematics and Engineering (SIAG/FME)\footnote{\url{http://wiki.siam.org/siag-fm/}} dedicated the Code Quest 2023 Programming Challenge to the implementation of a strategy to reduce the Conditional Value at Risk at level $\alpha$ ($CVaR_\alpha$), for a fixed degree of confidence $\alpha=0.9$, while keeping most of the return distribution mass above a certain threshold. That is, participants can control the initial (at time $t=0$) wealth distribution across the pools. This affects the return distribution at a fixed time $T$ in the future and, accordingly, the investment CVaR.\\
\linebreak
Minimizing the CVaR, seen as a portfolio optimization problem, is a well-known task in the financial literature. Among all, the cornerstone is the work by Rockafellar and Uryasev \cite{rockafellar2000optimization}. As it has been written in the framework of standard portfolio theory, it assumes linear dependency of the portfolio return (or loss) by its components' weights. However, this assumption fails in our context, where the relationship is well-known to be highly non-linear, as shown in the following section. Another popular approach is \cite{equiv_form}, where minimization is achieved by maximizing profit while constraining the CVaR$_{\alpha}$. Again, due to the complex nature of the return function in a CPMM context, the underlying assumptions of the approach are difficult to verify, and empirically this strategy does not bear fruit. Besides the classical theory of CVaR minimization, an alternative approach could be to see this problem as a standard optimization one and tackle it with the classical instruments of numerical analysis and optimization theory. Naive approaches, in this sense, obtain acceptable results in terms of accuracy but are very computationally expensive. Indeed, evaluating the map from the initial wealth distribution to the $CVaR_\alpha$ is extremely expensive, as we discuss later on. To mitigate these issues, we propose a three-step strategy that reduces the computational demand and provides accurate solutions. We first approximate the target function with a Kernel Ridge Regression (KRR). Then, we minimize the KRR approximation, and its minimum is used as the starting point for the last step, where the target function is directly minimized via Sequential Least Squares Programming (SLSQP).\\
\linebreak
The road map for this work is as follows: Section \ref{sec:cpmm} formalizes the problem by describing the mathematical aspects of the CPMM. Section \ref{sec:pro} provides a detailed overview of the Challenge task. Section \ref{sec:app} describes in detail our approach. In Section \ref{subsec:res}, the validity of our proposal is assessed both on the Challenge's framework and related settings. Finally, Section \ref{sec:conc} draws the conclusion.
\section{The Constant Product Market Maker}
\label{sec:cpmm}
\begin{figure}
    \centering
    \includegraphics[scale=0.45]{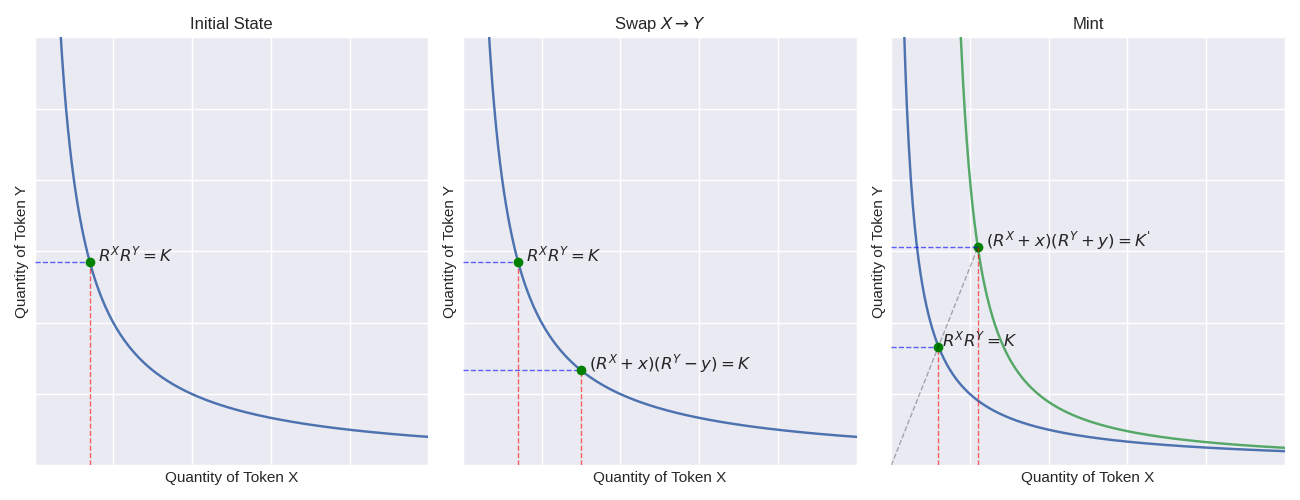}
    \caption{Visualization of the initial state, a swap from token X to token Y, and a mint operation. The burn operation resembles the mint operation, but the constant product rule shifts toward the axes rather than away from them in this scenario.}
    \label{fig:cpmm}
\end{figure}

This section aims to give an intuition about the functioning of CPMM and explain why traditional approaches to CVaR minimization cannot be applied in this context. As previously observed, CPMM takes its name from the fact that the product of coin reserves in the pool is kept constant. Therefore, assuming that the coins in the pool are X and Y, and the corresponding reserves are $R^X$ and $R^Y$, after any interaction with the pool we must have
\begin{equation}
    R^XR^Y=K \in \mathbb{R}
\label{cp_rule}
\end{equation}
Figure \ref{fig:cpmm} provides a graphical representation of the constant product rule, the swap $X \to Y$, and the mint operation.
\\
In the following, we explain in detail how the CPMM characterizes the basic operations of an AMM.
\subsection{Swap}
The swap is the action performed by a LT longing to exchange token X with Y, or the opposite. The CPMM relationship implies that if an LT swaps $x$ units of token X, the amount of token $y$ received is such that
\begin{equation}
\label{eq:0401_0000}
    R^X R^Y = \left( R^X+(1-\phi)x \right) \left( R^Y - y \right) \implies y = x \frac{(1-\phi)R^Y}{R^X + (1-\phi)x}
\end{equation}
where $\phi$ is the LT's fee to the pool. It is important to observe that the fee does not concur with the price formation but is entirely absorbed by the pool and distributed to the protocol and the LPs. In contrast to the Uniswap markets \cite{adams2020uniswap, adams2021uniswap}, in the Challenge's context the fee is added to the pool reserve of token X and, thus, it is implicitly used to remunerate LPs, as the value of the pool they own increases. This means that after the swap event, the reserves are updated as
\begin{equation}
    R^X \longrightarrow R^X+x \quad\text{and}\quad R^Y \longrightarrow R^Y-y
\end{equation}
In the case where the LT wants to swap $y$ units of Y against $x$ of X, the reasoning is absolutely the same
\begin{equation}
    R^X R^Y = \left( R^X-x \right) \left( R^Y + (1-\phi)y \right) \implies x = y \frac{(1-\phi)R^X}{R^Y + (1-\phi)y}
\end{equation}
Therefore, the reserves change according to
\begin{equation}
    R^X \longrightarrow R^X-x \quad\text{and}\quad R^Y \longrightarrow R^Y+y
\end{equation}
Lastly, it is possible to prove that the marginal price, defined as the limit $\displaystyle\lim_{y\rightarrow0}\frac{x}{y}$, is equal to the ratio between the reserves $\displaystyle\frac{R^X}{R^Y}$.\\
\subsection{Mint and burn}
From the Challenge point of view, it is more interesting to understand what happens to the LPs. They can add liquidity to the pool (the so-called mint operation) or withdraw tokens (burn). In the first case, they have to inject $x$ units of X and $y$ of Y so that the operation does not affect the marginal price. This means:
\begin{equation}
\label{eq:0401_0100}
    \frac{R^X}{R^Y} = \frac{R^X+x}{R^Y+y} \implies \frac{x}{y} = \frac{R^X}{R^Y}
\end{equation}
As a premium for the liquidity, the LP receives $l$ units of LP coins, where $l$ is determined as a function of the outstanding amount of LP coins (indicated with $L$):
\begin{equation}
l = L\frac{x}{R^X} = L\frac{y}{R^Y}
\end{equation}
The burn operation consists of giving back $l$ units of the LP coins to receive $x$ and $y$ units of tokens X and Y, according to the formula:
\begin{equation}
    x = \frac{l}{L}R^X \quad\quad and \quad\quad y = \frac{l}{L}R^Y
\end{equation}
From a real-world perspective, it is reasonable to assume the LPs only own $x$ units of token $X$. So, before the mint, they have to swap a certain amount $(1-\psi) x$ to obtain $y$ units of token Y, with $\psi\in(0,1)$. By substituting the quantity to swap $(1-\psi) x$ instead of $x$ in Equation \eqref{eq:0401_0000} and then $y$ in Equation \eqref{eq:0401_0100}, we obtain a second-order polynomial in $\psi$ whose positive root gives the amount to be swapped:
\begin{equation}
    \psi = 1 + \frac{(2-\phi)R^X}{2(1-\phi)x}\left( 1 - \sqrt{1 + 4\frac{x(1-\phi)}{R^X(2-\phi)^2}} \right)
\end{equation}
Then, when their investment period is over, they burn the tokens and swap their $y$ coins in the most advantageous pool. 
\section{The Challenge}
\label{sec:pro}
In the following, we work with multiple pools. Assuming $n$ pools, the notation previously introduced is adjusted accordingly. $\bm{R^X} = (R^X_1, \cdots, R^X_n)$ is the vector of coin X reserves in the pools. Similarly, we denote with $\bm{R^Y}$, $\bm{L}$, and $\bm{l}$ the vectors of coin Y reserves, the total amount of LP coins, and amount of LP coins held by a specific LP.\\
From the considerations in this subsection, it is now possible to understand the reason behind the highly non-linear dependence of the return from the initial wealth distribution. Assuming that after the LP provides liquidity, only swap operations occur, the portfolio weights $\bm{\theta}$ non-linearly affect the market state before the trades: $\bm{l}, \bm{R^X}, \bm{R^Y}, \bm{L}$. These impact the market state after the trade, which affects the conversion from $\bm{l}$ to $\bm{x}$ and $\bm{y}$ tokens when the investment is over, as well as the price for the LP to swap back $\bm{y}$ to $\bm{x}$.
\subsection{Formalization of the Minimization Problem}
The Challenge's task is to minimize the investment CVaR$_{\alpha}$ at a fixed level $\alpha=0.9$. Let's consider the total initial wealth of the LP as $x_0$, denominated in coin-X. The LP puts a fraction of his wealth on each of the $n$ pools. The portfolio weights are in the admissible set
\[
\mathcal{S} := \left\{ \bm{\theta}=(\theta_1\cdots\theta_n)\in[0,1]^n \ s.t. \ \sum_{j=1}^n\theta_j = 1 \right\}
\]
with $n=6$. As previously observed, the choice of the portfolio weights $\bm{\theta}$ affects both the number of LP coins generated, the pool state vector, and, as a consequence, the final (log) return of the investment $r_T=r_T(\bm{\theta})$. So far, the most used risk measure has been the Value at Risk (VaR), defined as the loss distribution quantile. That is:
\begin{equation}
    VaR_\alpha(r_T) = -\inf\left\{ z \in \mathbb{R} | \mathbb{P}(r_T \leq z) \geq 1-\alpha \right\}
\end{equation}
Nowadays, the attention is shifting from VaR to the CVaR, which is defined as the mean loss beyond the $\alpha$-quantile of the distribution, that is:
\begin{equation}
    \text{CVaR}_{\alpha}(r_T) = \frac{1}{1-\alpha}\int_{\alpha}^1 \text{VaR}_{s}(r_T)ds \approx \mathbb{E}[-r_T | -r_T \geq VaR_{\alpha}]
    \label{cvar}
\end{equation}
In order to stress the dependence of the objective function \eqref{cvar} from the weight vector $\bm{\theta}$ through the performance criterion $r_T$, we will write CVaR$_{\alpha}(\bm{\theta})$ instead of CVaR$_{\alpha}(r_T)$. Thus, for the Challenge to be completed, we have to find the optimal weights vector $\hat{\bm{\theta}}$ such that:
\begin{align}
    \hat{\bm{\theta}} \in \argmin_{\bm{\theta}\in\mathcal{S}} \text{CVaR}_\alpha(\bm{\theta})
\end{align}
subject to the following probabilistic constraint:
\begin{align}
\label{eq:constr_nl}
    \mathbb{P}\left[r_T > \xi\right] > q
\end{align}
with $\xi=0.05$ and $q=0.8$.\\
To summarize, we face a constrained minimization problem, with both equality and inequality constraints, linear and non-linear. However, we have empirically found that the non-linear probabilistic constraint in \eqref{eq:constr_nl} is automatically satisfied if we approximate the argmin for CVaR$_\alpha$. Thus, in the following, due to time and complexity reasons, we look at it only in the evaluation of the results and not even during the computation\footnote{We stress the fact that, in our framework, implementing the probabilistic constraint brings no conceptual or algorithmic difficulties.}.
\subsection{Computing the CVaR - The Simulation Process}
\label{sec:mi_serve}
\begin{figure}
    \centering
    \includegraphics[width=1.05\linewidth]{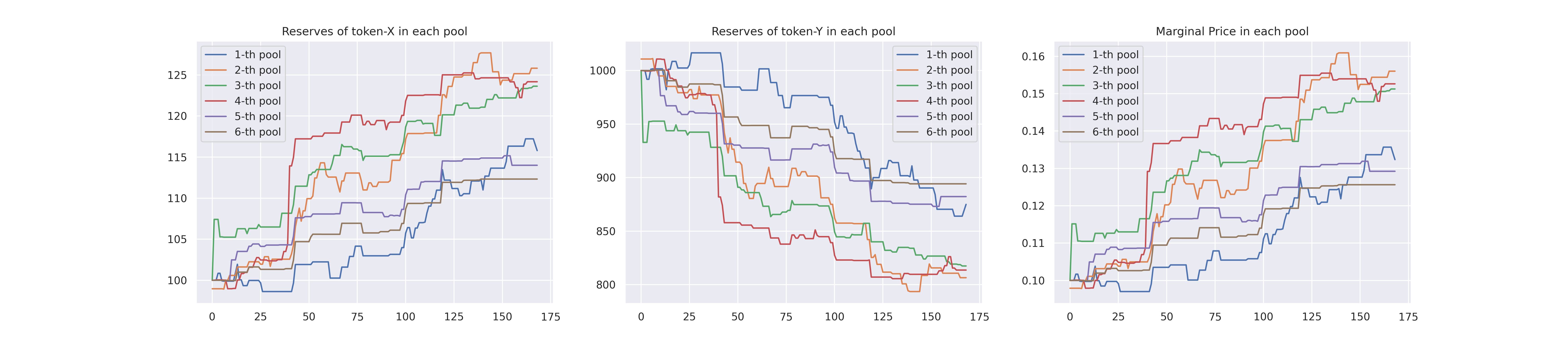}
    \caption{Example of a simulated path when using the Challenge parameters. Specifically, the left plot refers to the token X reserves $\bm{R^X}$; the central plot is for the coin Y reserves $\bm{R^Y}$; the right plot shows the marginal price.}
    \label{fig:sim_ex}
\end{figure}

Generally speaking, computing the CVaR of an unknown distribution is a hard task. However, for the purpose of the competition, the CVaR$_{\alpha}(\bm{\theta})$ function can be approximated by replacing the sample mean over the expectation in Equation \eqref{cvar}. Such a sample mean is computed over 1000 paths generated by a particular simulation engine provided by the Challenge's organizers. Specifically, the simulation process is based on a couple of assumptions. First, only swaps can occur in the pools. The direction of the swaps (from coin X to Y or the opposite) is randomly chosen without any dependence on the pool's state. Then, the arrivals' number is modeled via a Poisson distribution. Finally, the volumes of the swaps are sampled from a log-normal distribution.\\
In the following, when not otherwise specified, we work with $n+1$ dimensional vectors. The convention is that the first component is referred to a common event across all pools. Instead, when $j\ge1$, the $j$-th component is related to the $j$-th pool. The input parameter is a tuple $(\bm{\kappa}, \bm{p}, \bm{\sigma}, T, B)$ (whose value is provided by the organizer) such that:
\begin{itemize}
    \item $\bm{\kappa}=(\kappa_0,\cdots,\kappa_n)$ is the vector of arrival rates. In our case, $\bm{\kappa}=(0.25, 0.5,  0.5,  0.45,  0.45,  0.4,  0.3)$.
    \item $\bm{p}=(p_0, \cdots, p_m)$ represents the probability of swapping X to Y (as opposite of swapping Y to X). Observe that this probability changes according to the pool where the event takes place. We use $\bm{p}=(0.45, 0.45, 0.4, 0.38, 0.36, 0.34, 0.3)$.
    \item $\bm{\sigma}=(\sigma_0,\cdots,\sigma_n)$ represents the standard deviation for the log volume of the swap. Again, it changes with the pool. The default value in the competition is $\bm{\sigma}=(1, 0.3, 0.5, 1, 1.25, 2, 4)$.
    \item $T$ is the time horizon -that is, the length of the simulation. $T$ is fixed to 60.
    \item $B$ is the number of trajectories we want to simulate. In our case, $B=1000$.
\end{itemize}
From a programming perspective, the pools are described by a proper class, \texttt{pools}, that contains the attributes \texttt{Rx} and \texttt{Ry} describing the current reserves, and the swapping modules \texttt{swap\_x\_to\_y} and \texttt{swap\_y\_to\_x}. The pseudocode for the simulation process, viewed as a module of the class \texttt{pools}, is provided in Algorithm \ref{algo:sim}.
\begin{algorithm}[h]
\caption{Simulation routine provided by the Challenge's organizers}
\label{algo:sim}
\begin{algorithmic}[1] 
\State \textbf{Input}: \texttt{self}, $\bm{\kappa}, \bm{p}, \bm{\sigma}, T, B$.
\State \textbf{Output}: \texttt{pools}, \texttt{Rx\_t}, \texttt{Ry\_t}, \texttt{v\_t}, \texttt{event\_type\_t}, \texttt{event\_direction\_t}.
\State Initialize output lists.
\For{$k \gets 1 \text{ to } B$}
    \State Compute $K=\sum_{j=0}^n\kappa_j$ and draw $N \sim Poisson(KT)$.
    \State Initialize \texttt{curr\_pools} to \texttt{self}.
    \State Initialize \texttt{event\_type} and \texttt{event\_direction} to $N$-dimensional zeros vectors.
    \State Initialize \texttt{Rx}, \texttt{Ry}, and \texttt{v} to $N\times n$-dimensional zeros matrices.
    \For{$m \gets 1 \text{ to } N$}
        \State Draw the event type $j_m \sim Categorical(\frac{\kappa_0}{K},\cdots, \frac{\kappa_n}{K})$.
        \State Draw the event direction $d_m \sim Bernoulli(p_{j_m})$.
        \If {$d_m=0$}
            \State The drift vector $\bm{\mu}$ is initialized to the $n$-dimensional zero vector.
        \Else
            \State The drift vector $\bm{\mu}$ is initialized to $\log\frac{\text{\texttt{Rx}}}{\text{\texttt{Ry}}}$.
        \EndIf
        \If {$j_m=0$}
            \State The volume vector for the $j$-th pool is $\text{\texttt{v}[m,j]}\sim LogN(\mu_j, \sigma_j)$.
        \Else
            \State The volume vector for the $j_m$-th pool is $\text{\texttt{v}[m,}j_m\text{]}\sim LogN(\mu_{j_m}, \sigma_{j_m})$.
            \State The volume vector for the $j$-th pool, with $j\neq j_m$, is 0.
        \EndIf
        \If {$d_m=0$}
            \State Swap X to Y: \texttt{curr\_pools.swap\_x\_to\_y(v[m])}.
        \Else
            \State Swap Y to X: \texttt{curr\_pools.swap\_y\_to\_x(v[m])}.
        \EndIf

    \EndFor
    \State Add the new path features to the output lists.
\EndFor
\end{algorithmic}
\end{algorithm}\\
At the beginning of the simulation, we initialize the overall output list (line \textbf{3}). Then, we iterate over the paths' number. First, we generate the number of events according to a Poisson distribution with parameter equal to $T\sum_{j=0}^n\kappa_j$ (line \textbf{5}). Accordingly, we initialized the output for the specific trajectory (lines \textbf{6-8}). Then, we iterate among the events in the path. So, for the $m$-th event, the type $j_m$ is chosen (line \textbf{10}). It can be $j_m=0$ for a swap happening on all the pools or $1\le j_m\le n$ for a specific pool. The probability of each event is given by the corresponding $\kappa_j$ component, normalized by the sum of $\bm{\kappa}$ components. Then, the event direction $d_m$ is sampled according to a Bernoulli random variable with parameter $p_{j_m}$ (line \textbf{11}). The event volume is sampled from a log-normal distribution whose drift $\bm{\mu}=(\mu_1,\cdots,\mu_n)$ is a $n$ dimensional vector affected by $d_m$ (lines \textbf{12-16}):\begin{equation}
d_m=0 \implies \bm{\mu}=\bm{0} \quad\quad d_m=1 \implies \bm{\mu} = \log\frac{\text{\texttt{Rx}}}{\text{\texttt{Ry}}}
\end{equation}
Specifically, the volume is a $n$ dimensional vector whose components are all non-zeros if $j_m=0$. Otherwise, they are all zeros except one (lines \textbf{17-22}). In formula:
\begin{equation}
    j_m=0 \implies vol_j = LogN\left(\mu_j, \sigma_0\right) \forall j, \quad j_m\neq0 \implies vol_j = \begin{cases} LogN\left(\log\left( \mu_{j_m}, \sigma_{j_m}\right) \right) & if \ j=j_m \\ 0 & if \ j\neq j_m \end{cases}
\end{equation}
Finally, the volume is swapped (lines \textbf{23-27}). Once the simulation of a path is completed, the list containing all the generated samples is updated (line \textbf{29}). Observe that there are two types of outputs from the \texttt{simulate} module:
\begin{itemize}
    \item \texttt{pools}, \texttt{Rx\_t}, \texttt{Ry\_t}, \texttt{v\_t} depend on the initial pools state
    \item \texttt{event\_type\_t}, \texttt{event\_direction\_t} are independent on the initial state.
\end{itemize}
In the following, we exploit this difference to cut off the computational time. Indeed, during the CVaR minimization, we must perform the \texttt{simulate} module several times with different initial pools' states. Thus, our strategy is to generate the independent output once and then recycle it by adjusting the state-dependent variable.\\
Finally, an example of simulation is shown in Figure \ref{fig:sim_ex}.
\section{Our Approach}
\label{sec:app}
As previously discussed, in the context of this competition, there are several technical issues for CVaR$_\alpha$ minimization, specifically its non-linearity dependence from the input variable and the high estimation cost due to the simulation process. To tackle these problems, we design a three-step approach. We firstly approximate the target function for the CVaR$_\alpha$ using a Kernel Ridge Regression (KRR). Then, we minimize this (cheaper) function, and finally, we use the approximate solution for the wealth distribution as the starting point for the direct optimization of the CVaR$_\alpha$. We explain all the details in the next sections.
\subsection{Kernel Ridge Regression}
Kernel Ridge Regression \cite{vovk2013kernel} is a linear regression approach where the input variables are handled by a kernel $K$. Given a dataset $\left\{ \left(\bm{\theta}^{(i)},CVaR_\alpha^{(i)}\right) \right\}_{i=1}^N$, where each $\left(\bm{\theta}^{(i)},CVaR_\alpha^{(i)}\right)$ is a randomly generated pair of portfolio weights and CVaR$_\alpha$ value, the approximating function is:
\begin{equation}
f(\bm{\theta}; \bm{\alpha}) = \sum_{i=1}^N \alpha_i K(\bm{\theta}, \bm{\theta}^{(i)}) \quad\quad K(\bm{\theta}, \bm{\theta}^{(i)}) = -\sum_{j=1}^n \frac{\left(\theta_j - \theta^{(i)}_{j}\right)^2}{\theta_j + \theta^{(i)}_{j}}
\end{equation}
where we have used the additive $\chi^2$ kernel and $\bm{\alpha}=(\alpha_1,\cdots,\alpha_N)$ is the coefficients vector. The loss function is the same as in the ridge regression -that is, the mean square error with $L_2$ penalization on the coefficients. There is a closed form for fitting the model, so it requires a small computational time. We exploit the scikit learn implementation: \texttt{sklearn.kernel\_ridge.KernelRidge}.
\subsection{Sequential Least Square Programming}
The Sequential Least Squares Programming (SLSQP) algorithm is widely used to solve nonlinear optimization problems. It has been chosen due to its flexibility, as it can handle equality and inequality constraints and non-convex objective functions. SLSQP is a Sequential Quadratic Programming (SQP) method, where the minimum is found by iterations \cite{wright2006numerical}. At each step, the original problem is approximated by quadratic programming, which is solved via a quasi-Newton method. So, the formulation at step $k$ is:
\begin{equation}
\min_{\bm{d}} \ \left\{CVaR_\alpha(\bm{\theta}^{(k)}) + \nabla CVaR_\alpha(\bm{\theta}^{(k)})^T \bm{d} + \frac{1}{2}\bm{d}^T \nabla^2_{\theta\theta} \mathcal{L}\left(\bm{\theta}^{(k)}, \bm{\lambda}^{(k)}, \sigma^{(k)} \right) \bm{d}\right\}
\end{equation}
where $\mathcal{L}$ is the Lagrangian function associated to the problem and $\lambda$ and $\sigma$ are the Lagrange multipliers:
\begin{equation}
\mathcal{L}\left(\bm{\theta}, \bm{\lambda}, \sigma \right) = CVaR_\alpha(\bm{\theta}) - \bm{\lambda}^T \bm{\theta} - \sigma \left(1-\bm{1}^T\bm{\theta}\right)
\end{equation}
and the constraints can be reformulated as:
\begin{equation}
\bm{\theta}^{(k)} + I_n\bm{d} \ge \bm{0} \quad \& \quad \left(1-\bm{1}^T\bm{\theta}^{(k)}\right) - \bm{1}^T\bm{d} = 0 
\end{equation}
where $I_n$ is the identity matrix in $\mathbb{R}^{n\times n}$, and $\bm{0}, \bm{1}$ are the zero and one vectors in $\mathbb{R}^n$.
\subsection{The CVaR Minimization Approach}
\begin{figure}
    \centering
    \includegraphics[width=0.8\linewidth]{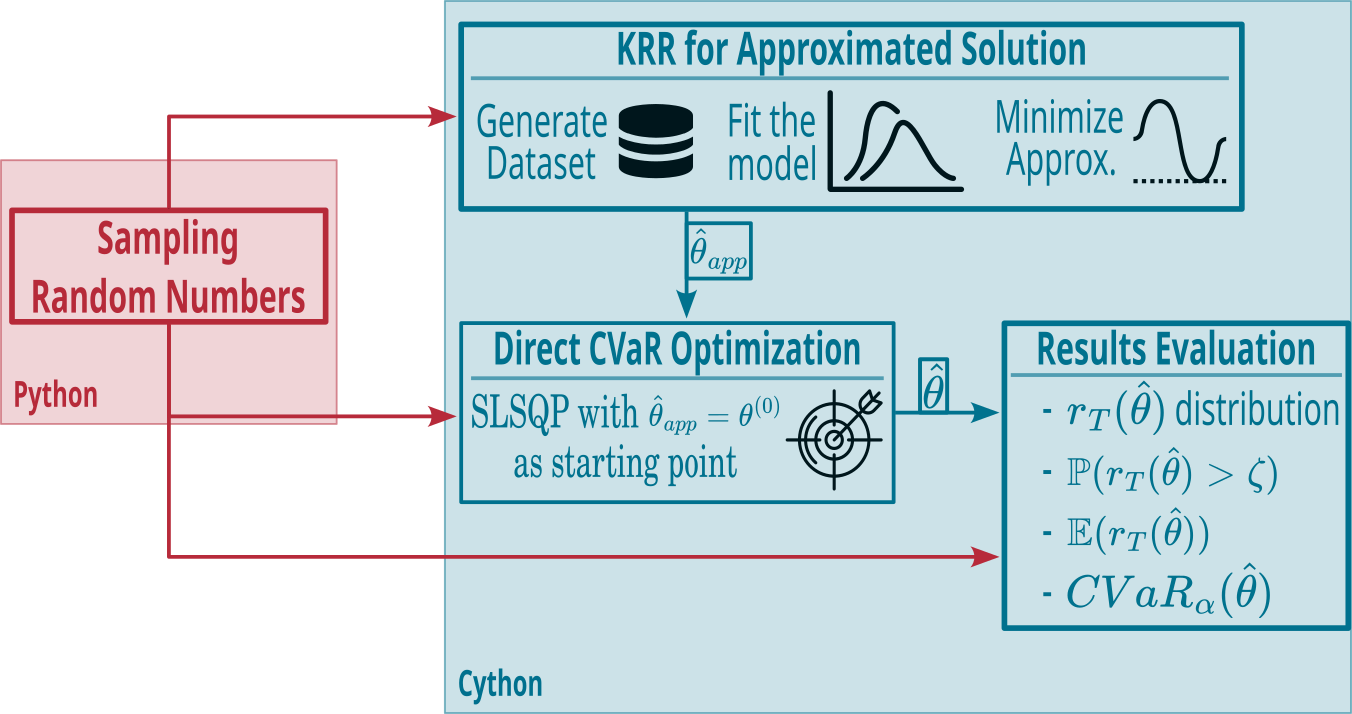}
    \caption{Graphical abstract describing our approach to the Challenge. The sampling of random numbers is the only part done in Python. Instead, the core of the approach is executed in Cython to reduce the computational time.}
    \label{fig:graph_abs}
\end{figure}

We propose a three-step approach for minimizing $CVaR_\alpha$ in the Challenge's framework. The strategy is summarized by the graphical abstract in Figure \ref{fig:graph_abs}
\\
Despite there is the expression \eqref{cvar} for the target function CVaR$_\alpha(\bm{\theta})$, its direct optimization would be extremely expensive, as every iteration is highly demanding. In this situation, we aim to achieve the minimization in more steps: firstly, we learn an approximation $f(\bm{\theta)}$ of the target function. Such an approximation has to be cheap to evaluate. Then, we minimize the approximating function. In this way, we aim to approach the minimum $\bm{\hat{\theta}}$ to obtain a good starting point for the last step, in which we use the SLSQP routine on expression \eqref{cvar}.\\
The first task is usually achieved with a neural network, such as in \cite{buchel2022deep}. However, the critical point is that generating the dataset for training the network could be as expensive as directly optimizing CVaR$_\alpha(\bm{\theta})$, leading to no significant advantages in computational time. So, we have used linear models, as, if well-specified, they require a small-size dataset to obtain a good fit. Specifically, the KRR has been used. The key point is correctly determining the dataset size $N$, as its generation can be extremely hard. We have found that even a small $N$ leads to accurate results with a very high $R^2$ determination coefficient. In particular, $N=10$ has been used. Thus, a dataset of observations $\left\{\left( \bm{\theta}^{(i)}, CVaR_\alpha^{(i)} \right)\right\}_{i=1}^N$ is created by randomly sampling  the $\bm{\theta}^{(i)}$ points. This dataset is used as the training set to fit the KRR model, that is, to find the coefficients vector $\bm{\hat{\gamma}}=(\gamma_1,\cdots,\gamma_N)$ such that:
\begin{equation}
\bm{\hat{\gamma}} = \argmin_{\bm{\gamma}} \sum_{i=1}^N \left( f(\bm{\theta}^{(i)}; \bm{\gamma}) - CVaR_\gamma^{(i)} \right)^2
\end{equation}
The goodness of the fit is proved by $R^2\approx0.995$. Finally, it is worth observing that the dataset creation can be easily parallelized to fully exploit the computational power and the processor's number of threads.\\
In the second step, the KRR is minimized by using the SLSQP to obtain an approximate solution to the Challenge's task:
\begin{equation}
\label{eq:0409_1200}
\bm{\hat{\theta}_{app}} = \argmin_{\bm{\theta}\in\mathcal{S}} f(\bm{\theta}; \bm{\hat{\gamma}})
\end{equation}
As the starting point, we use the equal-weighted portfolio. We explicitly point out that, due to the low KRR computational cost, this step requires just a fraction of a second. Furthermore, it is interesting to observe that the minimum CVaR$_{\alpha}$ in the KRR training set is -0.46\%, while CVaR$_{\alpha}(\bm{\hat{\theta}_{app}}) = -0.37\%$. This shows that KRR is really able to learn useful patterns and to efficiently approximate the CVaR$_{\alpha}$ function also in unseen points, and it is not merely copying the input data.\\
Last, we use the approximated optimal $\bm{\hat{\theta}_{app}}$ found in Equation \eqref{eq:0409_1200} as the starting point for directly minimizing the target function, CVaR$_\alpha(\bm{\theta})$, via SLSQP.
\subsection{Cython}
We propose introducing Cython in the optimization procedure to mitigate the complexity issue. Cython \cite{behnel2010cython} is a Python library that allows developers to write Python code that can be compiled into native C extensions. It combines the flexibility of the former with the efficiency of the latter. Cython is also compatible with most Python libraries and frameworks, making integrating with existing Python projects easy. The main reason for its performance is the usage of static type definitions for each variable. This is common in any compiled programming language like C++ and Fortran. Since the types of all variables are known at compile time, compilers can optimize the generated machine code more effectively. This leads to faster execution times and lower memory usage, as the compiler can make decisions about memory allocation and the best instructions to use based on the types of variables involved. Implementing Cython is a straightforward procedure summarized into Algorithm \ref{algo:cython_usage}. The \texttt{setup.py} file, needed for Cython to compile (line \textbf{4}), is a small Python script with compiling instructions. It can be found in \cite{source_code}.
\begin{algorithm}
\caption{Compiling Python code with Cython}
\label{algo:cython_usage}
\begin{algorithmic}[1]
\State Create a \texttt{.pyx} file.
\State Write the Python code in the \texttt{.pyx} file.
\State Add static type definitions for each variable.
\State Compile the \texttt{.pyx} code into a C extension using
\Statex \hspace{\algorithmicindent}\texttt{python setup.py build\_ext --inplace}
\State Import the compiled C extension in the Python script.
\end{algorithmic}
\end{algorithm}\\
Although Cython allows for an important speed-up in code execution, in the context of the Challenge, it presents the drawback of handling the internal numpy random number generation differently from Python. This can bias the project evaluation due to the different evolution scenarios from the competing teams. To solve this issue and further reduce the simulations' complexity, we generate all the random numbers simultaneously before using Cython. That is, we generate and store the arrays used to determine the orders arrivals, types, and directions -that is, the arrays \texttt{event\_type\_t} and \texttt{event\_direction\_t} described in Section \ref{sec:mi_serve}. Then, when using Cython, we load these arrays. This procedure can drop the computational time by almost 20\%.
\section{Experimental Results}
\label{subsec:res}
This section compares our team's proposal, QuantHub, with those of competing teams. First, we examine the performance in the Challenge's environment. Later, we vary the simulation parameters to study the robustness of the proposed approaches.
\subsection{Challenge's Result}
First, we study the performance obtained by our algorithm in the Challenge's environment. We use the same market parameters and random seed provided by the quest organizers, and we evaluate the approach with the same metrics they consider. Furthermore, we compare QuantHub results with those obtained by Finatics \cite{finatics} and Elagnitram \cite{elagnitram}. To our knowledge, they are the only teams that post their code publicly. Furthermore, in this subsection, we also show the result by Blanco, a finalist team that kindly shared their final report. Unfortunately, as we do not have their exact code, the comparison with Blanco is limited to this stage, where we use the value of the metrics declared in their report. Finally, the approaches followed by our competitor are discussed in detail in Appendix \ref{app:competing}.\\
\linebreak
As already mentioned before, the parameters provided by the organizers are the following:
\begin{equation}
    \begin{array}{rc} \bm{\kappa} = & (0.25, 0.5,  0.5,  0.45,  0.45,  0.4,  0.3) \\ \bm{p} = & (0.45, 0.45, 0.4, 0.38, 0.36, 0.34, 0.3) \\ \bm{\sigma} = & (1, 0.3, 0.5, 1, 1.25, 2, 4) \\ Random \ seed = & 4294967143 \end{array}
\end{equation}
The results obtained by the different competitors are summarized in Table \ref{tab:challenge_results}.
\begin{table}[h!]
\centering
\begin{tabular}{|c||c|c|c|}
\hline
\rowcolor{gray!15}
& $\bm{\hat{\theta}}$ & $\mathbb{P}\left[r_T > \xi\right]$ & CVaR$_{\alpha}$ \\
\hline
\hline
\textbf{Blanco} & $[0.1308, \ 0.2480, \ 0.2209, \ 0.1438, \ 0.2394, \ 0.0173]$ & 0.847 & -0.376\% \\
\textbf{Finatics} & $[0.1597, \ 0.3029, \ 0.1901, \ 0.1168, \ 0.2304, \ 0.0000]$  & 0.847 & -0.364\% \\
\textbf{Elagnitram} & $[0.1749, \ 0.1643, \ 0.1498, \ 0.1822, \ 0.2272, \ 0.1016]$ & 0.840 & -0.460\% \\
\textbf{QuantHub} & $[0.1285, \ 0.2956, \ 0.1897, \ 0.1568, \ 0.2294, \ 0.0000]$ & 0.846 & -0.3627\% \\
\hline
\end{tabular}
\caption{Summary of Challenge's results. The table shows both the optimal solution $\bm{\hat{\theta}}$, the value of the constraint $\mathbb{P}[r_T>\xi]$, and the CVaR$_\alpha$. Our proposal, \textbf{QuantHub}, is compared with the finalist \textbf{Blanco} (which kindly shared their report with us) and with the approaches by the teams \textbf{Finatics} and \textbf{Elagnitram} (the only publicly available on GitHub).}
\label{tab:challenge_results}
\end{table}\\
It is worth noting that our strategy and the one employed by Finatics prioritize distributing the least wealth to the last pool. The difference with the other pools is of different orders of magnitude. As discussed in the Finatics final report, this behavior aligns with the findings of Cartea et al. \cite{cartea} regarding predictable losses (that is, the negative component of an LP return) in a liquidity provision strategy within the context of CPMM. Assuming a geometric Brownian motion for the marginal price, Cartea et al. demonstrated that these predictable losses are directly proportional to the standard deviation of the price. Indeed, the sixth pool exhibits the highest volatility, as evident from the $\bm{\sigma}$ parameter. However, it should be pointed out that even though the volatility of the fifth pool is high, the amount of capital allocated to it is the second. To clarify this aspect, we replicated the same experiment by putting the volatility of the fifth pool equal to 4, as the last pool. We find a clear negative dependence of the optimal weight of a pool from its variance, with a diminution in the fifth pool's allocation of about 35\%. Indeed, the optimal theta found by QuantHub is:
\[
\bm{\hat{\theta}} = [0.124, 0.441, 0.216, 0.148, 0.0]
\]
Finally, to assess the validity of the QuantHub double optimization procedure (targeted on both the approximating function and the original one), we carry out an ablation study. That is, we compare the QuantHub approach with its single steps using different metrics. In Table \ref{tab:res}, we show the results obtained by only the KRR component in the first two steps and the SLSQP in the third step -that is, with $\bm{\theta}^{(0)}$ initialized to the equal-weighted portfolio. To provide a wider picture, the randomized Grid Search (carried out by iterating over 10000 random starting points) is also considered.
\begin{table}[h!]
\centering
\begin{tabular}{|c|c|c|c|c|}
\hline
\rowcolor{gray!15}
 & \textbf{Grid Search} & \textbf{KRR} & \textbf{SLSQP} & \textbf{QuantHub}  \\ \hline
$\bm{\mathbb{P}\left[r_T > \xi\right]}$ & 84.50\% & 84.20\% & 84.50\% & \textbf{84.60\%} \\ \hline
$\bm{\mathbb{E}[r_T]}$ & 16.0985\% & \textbf{16.1744\%} & 16.1306\% & 16.1250\% \\ \hline
VaR$_{\alpha}$ & 3.1832\% & 3.2228\% & 3.2196\% & \textbf{3.2237\%} \\ \hline
CVaR$_{\alpha}$ & -0.3693\% & -0.3731\% & -0.3632\% & \textbf{-0.3627\%} \\ \hline
$\hat{\bm{\theta}}$ & \begin{tabular}[c]{@{}c@{}}[0.11057, 0.34389,\\ 0.17021, 0.13990,\\ 0.22973, 0.00567]\end{tabular}  & \begin{tabular}[c]{@{}c@{}}[0.14408, 0.29763,\\ 0.17649, 0.16160,\\ 0.19700, 0.02320]\end{tabular}  & \begin{tabular}[c]{@{}c@{}}[0.12360, 0.29529,\\ 0.19724, 0.15761,\\ 0.2262, 0.00001]\end{tabular} & \begin{tabular}[c]{@{}c@{}}\textbf{[0.12848, 0.29556,}\\ \textbf{0.18971, 0.15680,}\\ \textbf{0.22943, 0.00001]}\end{tabular} \\ \hline
\end{tabular}
\caption{Comparison of QuantHub approach with its single steps. The whole procedure is compared with the first two steps (\textbf{KRR}) and the last one (\textbf{SLSQP}). Furthermore, to understand the goodness of our proposal, we add the randomized Grid Search (\textbf{Grid Search}) to the comparison.}
\label{tab:res}
\end{table}\\
As shown, the optimal vectors $\hat{\bm{\theta}}$ obtained by all the methods are very similar. However, our strategy slightly outperforms the competitors in terms of the performance metric. In particular, the KRR performs worse, and the optimal $CVaR_\alpha$ it finds is 2.7\% lower than the QuantHub one. This shows the utility of the SLSQP step in our procedure: without it, the computational times would be lower, but the solution accuracy would be damaged.
\subsection{Stability Experiments}
To assess the goodness of our proposal, we have compared it with the competing approaches described in the previous subsection by letting the random seed vary. Specifically, 100 different random seeds have been used. It is worth noting that the Elagnitram approach sometimes does not reach convergence. The comparison results are shown in Table \ref{tab:diff_seeds}. The table reports the mean and standard deviation for both the probabilistic constraint, the mean return, VaR$_\alpha$, CVaR$_\alpha$
\begin{table}
    \centering
\begin{tabular}{|c|c|c|c|}
\hline
\rowcolor{gray!15}
 & \textbf{QuantHub} & \textbf{Finatics} & \textbf{Elagnitram}  \\ \hline
$\bm{\mathbb{P}\left[r_T > \xi\right]}$ (\%) & 81.15 $\pm$ 1.02 & 81.17 $\pm$ 1.01 & 81.5 $\pm$ 0.91 \\ \hline
$\bm{\mathbb{E}[r_T]}$ & 0.1686 $\pm$ 0.0114 & 0.1686 $\pm$ 0.0114 & 0.1691 $\pm$ 0.0105 \\ \hline
VaR$\bm{_{\alpha}}$ & 0.021 $\pm$ 0.0025 & 0.0209 $\pm$ 0.0024 & 0.0197 $\pm$ 0.0029 \\ \hline
CVaR$\bm{_{\alpha}}$ & -0.0145 $\pm$ 0.0029 & -0.0145 $\pm$ 0.0029 & -0.0152 $\pm$ 0.0035 \\ \hline
Times (s) & 65.9 $\pm$ 18.7 & 1053.4 $\pm$ 23.4 & 155.1 $\pm$ 62.7 \\ \hline

\end{tabular}
\caption{Comparison between QuantHub, Finatics, and Elagnitram approaches. The algorithms run over 100 random seeds (from 1 to 100). The mean and the standard deviation are shown. The Elagnitram algorithm does not return any output on about half of the experiments.}
\label{tab:diff_seeds}
\end{table}\\
The experimental result shows that QuantHub and Finatics obtain almost the same accuracy. The main difference is in computational time, where QuantHub is more than 15 times faster than its competitor. Instead, compared with Elagnitram, we find the latter is slightly faster, but the QuantHub CVaR results are better than the Elagnitram ones. The mean gain is more than 4.5\%. In general, Elagnitram optimal solutions seem to perform better in the return mean (the first two rows in the table) but lack accuracy when looking at the risk measures (the third and the fourth rows), which are the true targets of the competition.
\subsection{Insights into Generalization Property}
A more detailed comparison between the three algorithms is done by changing the investment parameters. The aim is to investigate the strategies' robustness and ensure they do not just overfit the parameters the Challenge's organizers provided. To achieve this goal, we allow both the temporal horizon $T$ and the confidence level $\alpha$ to vary. A total of 10 experiments are carried out. Five use the same competition parameters except for $T$ that varies in $[40, 50, 60, 70, 80]$. The other experiments use the same parameters and $\alpha\in[0.85, 0.875, 0.9, 0.925, 0.95]$. Each model is run 50 times for every experiment with different random seeds. The results are summarized in Table \ref{tab:T_alpha_comp}. We show the optimal CVaR$_\alpha$, the value of the probabilistic constraint, and the computational time in terms of mean and standard deviation across the different seeds.

\begin{table}[h]
    \centering
    \begin{adjustwidth}{-1.1cm}{}
\begin{tabular}{|c|c||c|c|c|c|c|}
\hline
\rowcolor{gray!15}
 & $T$= & 40 & 50 & 60 & 70 & 80\\
\hline
\multirow{3}{*}{\textbf{QuantHub}} & CVaR$_\alpha$\%$\cdot10^{-1}$ & $-0.367 \pm 0.038$ & $-0.266 \pm 0.029$ & $-0.153 \pm 0.022$ & $-0.019 \pm 0.032$ & $0.103 \pm 0.043$ \\
 & $\mathbb{P}[r_T > \xi]$\% & $63.18 \pm 48.232$ & $74.22 \pm 43.742$ & $80.7 \pm 39.465$ & $86.7 \pm 33.957$ & $90.1 \pm 29.866$ \\
 & time (s) & $178.0 \pm 56.0$ & $208.0 \pm 64.0$ & $234.0 \pm 86.0$ & $247.0 \pm 94.0$ & $355.0 \pm 120.0$ \\

\hline
\multirow{3}{*}{\textbf{Finatics}} & CVaR$_\alpha$\%$\cdot10^{-1}$ & $-0.367 \pm 0.038$ & $-0.266 \pm 0.029$ & $-0.153 \pm 0.022$ & $-0.019 \pm 0.031$ & $0.103 \pm 0.043$ \\
 & $\mathbb{P}[r_T > \xi]$\% & $63.06 \pm 48.264$ & $74.26 \pm 43.72$ & $80.7 \pm 39.465$ & $86.78 \pm 33.871$ & $90.08 \pm 29.893$ \\
 & time (s) & $2250.0 \pm 341.0$ & $4416.0 \pm 683.0$ & $4869.0 \pm 1131.0$ & $5149.0 \pm 1513.0$ & $5936.0 \pm 1674.0$ \\

\hline
\multirow{3}{*}{\textbf{Elagnitram}} & CVaR$_\alpha$\%$\cdot10^{-1}$ & nan & nan & $-0.16 \pm 0.031$ & $-0.028 \pm 0.032$ & $0.094 \pm 0.042$ \\
 & $\mathbb{P}[r_T > \xi]$\% & nan & nan & $81.75 \pm 38.626$ & $86.56 \pm 34.108$ & $89.94 \pm 30.08$ \\
 & time (s) & nan & nan & $303.0 \pm 67.0$ & $265.0 \pm 85.0$ & $621.0 \pm 651.0$ \\
\hline
\hline
\rowcolor{gray!15}
 & $\alpha$= & 0.85 & 0.875 & 0.9 & 0.925 & 0.95 \\
\hline
\hline
\multirow{3}{*}{\textbf{QuantHub}} & CVaR$_\alpha$\%$\cdot10^{-1}$ & $-0.154 \pm 0.022$ & $-0.153 \pm 0.022$ & $-0.153 \pm 0.022$ & $-0.154 \pm 0.022$ & $-0.155 \pm 0.022$ \\
 & $\mathbb{P}[r_T > \xi]$\% & $80.7 \pm 39.465$ & $80.7 \pm 39.465$ & $80.7 \pm 39.465$ & $80.64 \pm 39.512$ & $80.62 \pm 39.527$ \\
 & time (s) & $50.0 \pm 12.0$ & $47.0 \pm 13.0$ & $62.0 \pm 19.0$ & $64.0 \pm 22.0$ & $79.0 \pm 31.0$ \\

\hline
\multirow{3}{*}{\textbf{Finatics}} & CVaR$_\alpha$\%$\cdot10^{-1}$ & $-0.154 \pm 0.022$ & $-0.153 \pm 0.022$ & $-0.153 \pm 0.022$ & $-0.154 \pm 0.022$ & $-0.155 \pm 0.022$ \\
 & $\mathbb{P}[r_T > \xi]$\% & $80.68 \pm 39.481$ & $80.74 \pm 39.434$ & $80.72 \pm 39.45$ & $80.72 \pm 39.45$ & $80.58 \pm 39.558$ \\
 & time (s) & $1144.0 \pm 183.0$ & $977.0 \pm 10.0$ & $970.0 \pm 1.0$ & $982.0 \pm 9.0$ & $983.0 \pm 7.0$ \\

\hline
\multirow{3}{*}{\textbf{Elagnitram}} & CVaR$_\alpha$\%$\cdot10^{-1}$ & $-0.162 \pm 0.026$ & $-0.162 \pm 0.026$ & $-0.162 \pm 0.026$ & $-0.162 \pm 0.027$ & $-0.161 \pm 0.026$ \\
 & $\mathbb{P}[r_T > \xi]$\% & $81.133 \pm 39.124$ & $81.2 \pm 39.071$ & $81.2 \pm 39.071$ & $81.233 \pm 39.045$ & $81.167 \pm 39.098$ \\
 & time (s) & $49.0 \pm 14.0$ & $49.0 \pm 12.0$ & $49.0 \pm 13.0$ & $74.0 \pm 36.0$ & $56.0 \pm 24.0$ \\
\hline
\end{tabular}
    \end{adjustwidth}
    \caption{Comparison between different approaches as the time horizon $T$ and the confidence level $\alpha$ vary. The parameters are the same as in the competition, except for the one given in the column header. We run the algorithms 50 times for each experiment with different random seeds. The table shows the mean and the standard deviation obtained for the optimal CVaR$_\alpha$ value, the probabilistic constraint, and the computational time.}
    \label{tab:T_alpha_comp}
\end{table}
Some considerations arise from the behavior as $T$ varies. First, the results obtained by QuantHub and Finatics are very close to each other regarding the CVaR. The only noticeable difference is in the computational time. Then, the optimal CVaR increases almost linearly with the time horizon. Indeed, by the construction of the market, there is a positive drift in the LP's wealth process, given by the fees paid by the LTs. Furthermore, the probability of returns bigger than 5\% increases as $T$, even if slower. Finally, there is an increase in the computational time, given by the larger amount of events that need to be simulated. However, QuantHub and Elagnitram show an abrupt increase when $T$ shifts from 70 to 80, even if their time is far below the Finatics one.\\
As for the results as a function for $\alpha$, it holds the same considerations about the comparison between QuantHub and Finatics. That is, their CVaR is very similar, but the former is far faster than the latter. Finally, there are no noticeable differences in the optimal CVaR value, nor particular patterns in the computational time.
\section{Conclusion}
\label{sec:conc}
This work presents our approach to the SIAG/FME Code Quest 2023. The Challenge requires designing a static investment strategy that minimizes the CVaR when providing liquidity to multiple pools in a Decentralized Exchange. Specifically, the participants were provided with a simulation engine that mimics the arrival of orders in the market. By choosing the initial wealth distribution, agents modify the final returns distribution and, accordingly, the investment CVaR. Our approach to the Challenge is based on three steps. The first mitigates the evaluation complexity by approximating the target function with KRR. Then, the KRR is minimized to obtain a suitable starting point for the third step, where SLSQP is used to refine the approximated solution. An ablation study is carried out to assess the validity of all the steps involved. Furthermore, the comparison with the Grid Search proves the quality of the optimal point found.\\
\linebreak
In the experimental stage, our proposal is compared with two algorithms designed for the same competition. We compare the approaches according to the level of the ultimate CVaR and the computational time. Furthermore, to assess the robustness, we let the competition parameters vary. The overall results show the efficiency of our proposal, which can achieve a better CVaR than competitors while keeping the computational time low.\\
The main limitation of this work is related to the simulation engine used in the Challenge. Indeed, it describes a simplified version of a Decentralized Market, where no attention is paid to avoiding arbitrage opportunities, and the events do not depend on the state of the pools. So, in the future, it can be interesting to exploit the same approach in different contexts -e.g., in the Uniswap markets. Furthermore, the simulation engine should be better designed to avoid the abovementioned issues. This task could be achieved either with statistical models or a generative approach.\\
\linebreak
Finally, it should be noted we have worked in a simplified environment. Further enhancement could include considering more sophisticated markets, such as the Concentrated Liquidity of Uniswap v3 \cite{adams2021uniswap}, or the LP problem in its whole complexity -that is, by also taking into account the so-called Impermanent Loss \cite{heimbach2022risks}.
\section*{Acknowledgment}
We would like to thank the Quest Committee for the extraordinary opportunity they gave us. We would also like to thank Gianluca Palmari for his valuable suggestions, as well as our professors, Fabrizio Lillo and Piero Mazzarisi, for helping us in our work.
\appendix
\section{Appendix - Competing Approaches}
\label{app:competing}
In this appendix, we describe some competing approaches to the Challenge. Firstly, we discuss the proposal of the team Blanco, which has won the front-running prize. Then, we show the methodologies from Finatics and Elagnitram, which have been publicly shared on GitHub.
\subsection{Blanco}
The pseudocode is presented in Algorithm \ref{algo:competing_algo}, where the \texttt{generate\_market} function simulates the market paths. The idea is to find the optimal $\bm{\hat{\theta}}$ iteratively. Each step generates one thousand market paths, $\bm{r}_T=\{r_T^{(j)}\}_{j=1}^B$ and the approximation is updated via gradient descent.
\begin{algorithm}
\caption{Blanco's Algorithm}
\label{algo:competing_algo}
\begin{algorithmic}[1] 
\State \textbf{Data}: $N_{\text{pools}}$, $N_{\text{batch}}$, $R^{X}$, $R^{Y}$, $\alpha$, $\gamma$, $q$, $\zeta$, $\kappa$, $\sigma$, $\bm{\theta}$, $\omega$, $\beta$, $N_{\text{iter}}$, $N_{\text{GD}}$.
\State \textbf{Result}: Optimal weights.
\State $w_1 := w$
\For{$i \gets 1 \text{ to } N_{\text{iter}}$}
    \State $(\tilde{x}, \tilde{y}, R^{X}, R^{Y}, \phi) := \text{\texttt{generate\_market}}(w_i)$
    \State $w_1 = w_i$
    \For{$j \gets 1 \text{ to } N_{\text{GD}}$}
        \State $\bar{x}_{\text{burn}} := \bar{x} \left(\tilde{\theta}_j \odot \theta_{i}^{-1} \right)$
        \State $\bar{y}_{\text{burn}} := \bar{y} \left(\tilde{\theta}_j \odot \theta_{i}^{-1} \right)$
        \State $\bar{x}_{\text{swap}} := \frac{\bar{y}_{\text{burn}} \odot (1-\phi) \odot R^{X}}{R^{Y} + (1-\phi) \odot \bar{y}_{\text{burn}}}$
        \State $\bar{r} := \log(\bar{x}_{\text{burn}} + \bar{x}_{\text{swap}}) - \log(x_0)$
        \State Compute the loss $\ell(w_j, r; j) := \omega_1 \ell_1 + \omega_2 \ell_2 + \omega_3 \ell_3 + \omega_4 \ell_4$
        \State Compute gradients $\nabla \tilde{\theta}_j$ with respect to the loss $\ell$.
        \State $\tilde{\theta}_{j+1} := \tilde{\theta}_j - \beta \cdot \nabla \tilde{\theta}_j$
    \EndFor
    \State $w_{i+1} := w_{N_{\text{GD}}+1}$
\EndFor
\State \Return $w_{N_{\text{iter}}+1}$
\end{algorithmic}
\end{algorithm}\\
As we can see from the Algorithm \ref{algo:competing_algo}, at each iteration of the nested \texttt{for} loop, the quantities of tokens X and Y to burn are reweighted according to the current gradient descend weight $\bm{\theta}$ (lines 8 and 9). Then, the total loss is computed according to
\begin{equation}
\ell(\bm{\theta}, \bm{r}) = \omega_1 l_1 + \omega_2 l_2 + \omega_3 l_3 + \omega_4 l_4
\end{equation}
where 
\begin{equation}
    \begin{array}{cc}\ell_1 = \ell_1(\bm{\theta}, \bm{r}_T) = &\ \displaystyle \left( q_\alpha - \frac{1}{B} \sum_{i=1}^{B} G(1000 \cdot (r_T^{(i)} - \zeta)) \right)^2_+ \\
    \ell_2 = \ell_2(\bm{\theta}, \bm{r}_T) = &\ \displaystyle \frac{1}{n} \sum_{i=1}^{n} (-\theta_i)_+ \\
    \ell_3 = \ell_3(\bm{\theta}, \bm{r}_T) = &\ \displaystyle \left( \sum_{i=1}^{B} \theta_i - 1 \right)^2 \\
    \ell_4 = \ell_4(\bm{\theta}, \bm{r}_T) = &\ \displaystyle \text{CVaR}_{\alpha} = \frac{\sum_{i=1}^{B} (-r_T^{(i)}) \mathbf{1}_{\{-r_i > q_{\alpha}(-\bm{r}_T)\}}}{|\{ i \in \{1, \ldots, B\} : -r_T^{(i)} > q_{\alpha}(-\bm{r}_T) \}|}\end{array}
\end{equation}
The $\omega_i$ are scalar quantities set to have all the $\ell_i$ of the same order of magnitude. $G(x)=(1+e^{-x})$ is the sigmoid function, $(x)_+$ is the ReLU function and $q_{\alpha}$ is the $\alpha$-quantile function. \\
The initial value of the weights was chosen according to the profitability of a strategy in which all the wealth is provided to a single pool. Then, they ranked the best pools from the worst to the best and set the initial weights evenly distributed between the first three best pools. Convergence analysis concerning different random seeds and initial weights proved that this choice led to the best results.
\subsection{Finatics}
Finatics used a Stochastic Gradient Descend approach to search for an optimal value of the weight $\bf{\theta}$. In their report \cite{finatics}, they present a classic Stochastic Gradient Descend where the constraints on the weights $\theta_i$ are enforced via a projection onto the admissible set $\mathcal{S}$ where $\sum_i \theta_i = 1$ and $\theta_i>0$ $\forall i$. Furthermore, they do not actively enforce the probability constraint on the expected return since it is never unsatisfied. This is in accordance with our findings. The Finatics approach method is presented in Algorithm \ref{algo:finatics_sdg}.
\begin{algorithm}
\caption{Finatics' Algorithm}
\label{algo:finatics_sdg}
\begin{algorithmic}[1]
\Require Number of iterations $N_{iter}$.
\State Initialize randomly $\bm{\theta}^{(1)} = (\theta_1, \dots, \theta_6) \in \mathcal{K}$
\State Set $\eta$ to a chosen learning rate value.
\For{$n = 1, \dots, N_{iter}$}
    \State Simulate 1000 returns $\{r^{(1)}_T, \dots, r_T^{(1000)}\}$ based on $\bm{\theta}_n$ and calculate CVaR
    \State Compute $g_n = \nabla_{\bm{\theta}} CVaR(\bm{\theta}_n)$.
    \State Update $\bm{\theta}_{n+1} = \text{proj}_{\mathcal{S}}(\bm{\theta}_n - \eta g_n)$.
\EndFor
\State \textbf{return} $\theta_{N_{iter}}$
\State Select $\bm{\theta}$ to minimize CVaR that meets probability constraint
\end{algorithmic}
\end{algorithm}
\subsection{Elagnitram}
Elagnitram devised an optimization procedure based on the Gradient Descend algorithm. Due to the constraint $\sum_i\theta_i=1$, they optimized only five out of six weights, computing the last one thanks to the above relation. A great effort is put into ensuring the probabilistic constraint $\mathbb{P}[r_T>\xi]>q$. In order to guide the optimization in the regions where the probability constraint is satisfied, they defined two constants, namely $\delta_1$ and $\delta_2$, with $\delta_1<\delta_2$. Let $\psi$ be the $(1-q)$-quantile of the returns $r_T$. Then:
\begin{enumerate}
\item If $\psi<\zeta + \delta_1$, they assume the probability constrained violated, since to obtain a cumulative return probability greater than $q$, one should start counting from a $r_T<\zeta$.
\item If $\psi>\zeta + \delta_2$, the probability constraint is satisfied, and they update the weights using the gradient of the CVaR, assuming that the updated weights would not violate the constraint.
\item If $\zeta + \delta_1 < \psi <\zeta + \delta_1$, the current weights satisfy the probability constraint, but they should change in such a way that the $(1-q)$-quantile does not decrease too much. In order to do so, they find the suitable search direction for the updating perpendicular to the gradient of $\xi$.
\end{enumerate}
Additionally, they were able to use an approximated relationship for the pools' evolution that lightened the computational hurdle. Finally, they check at each optimization step if all the weights are positive and if they sum to 1. If not, they search for the largest admissible step size that satisfies these constraints. Further details can be found on their report in \cite{elagnitram}.
\bibliographystyle{abbrv}
\bibliography{main}

\begin{thebibliography}{10}

\bibitem{adams2020uniswap}
H.~Adams, N.~Zinsmeister, and D.~Robinson.
\newblock Uniswap v2 core, 2020.
\newblock {\em URL: https://uniswap. org/whitepaper. pdf}, 2020.

\bibitem{adams2021uniswap}
H.~Adams, N.~Zinsmeister, M.~Salem, R.~Keefer, and D.~Robinson.
\newblock Uniswap v3 core.
\newblock {\em Tech. rep., Uniswap, Tech. Rep.}, 2021.

\bibitem{alotaibi2021smart}
L.~S. Alotaibi and S.~S. Alshamrani.
\newblock Smart contract: Security and privacy.
\newblock {\em Computer Systems Science \& Engineering}, 38(1), 2021.

\bibitem{behnel2010cython}
S.~Behnel, R.~Bradshaw, C.~Citro, L.~Dalcin, D.~S. Seljebotn, and K.~Smith.
\newblock Cython: The best of both worlds.
\newblock {\em Computing in Science \& Engineering}, 13(2):31--39, 2010.

\bibitem{buchel2022deep}
P.~B{\"u}chel, M.~Kratochwil, M.~Nagl, and D.~R{\"o}sch.
\newblock Deep calibration of financial models: turning theory into practice.
\newblock {\em Review of Derivatives Research}, pages 1--28, 2022.

\bibitem{cartea2023decentralised}
{\'A}.~Cartea, F.~Drissi, and M.~Monga.
\newblock Decentralised finance and automated market making: Predictable loss and optimal liquidity provision.
\newblock {\em arXiv preprint arXiv:2309.08431}, 2023.

\bibitem{cartea}
{\'A}.~Cartea, F.~Drissi, and M.~Monga.
\newblock Predictable losses of liquidity provision in constant function markets and concentrated liquidity markets.
\newblock {\em Applied Mathematical Finance}, 30(2):69--93, 2023.

\bibitem{defillama}
DefiLlama.
\newblock \url{https://defillama.com}.
\newblock 2023.

\bibitem{source_code}
D.~M. Di~Nosse and F.~Gatta.
\newblock {QuantHub Code Repository}.
\newblock \url{https://github.com/DanieleMDiNosse/SIAM_code_challange/tree/main}, 2024.

\bibitem{elagnitram}
elagnitram.
\newblock {elagnitram Code Repository}.
\newblock \url{https://github.com/Jue-Edin/elagnitram}, 2024.

\bibitem{finatics}
T.~N. Hai and L.~N. Tuan.
\newblock {Finatics Code Repository}.
\newblock \url{https://github.com/hitwooo/siag-fme24}, 2024.

\bibitem{heimbach2022risks}
L.~Heimbach, E.~Schertenleib, and R.~Wattenhofer.
\newblock Risks and returns of uniswap v3 liquidity providers.
\newblock In {\em Proceedings of the 4th ACM Conference on Advances in Financial Technologies}, pages 89--101, 2022.

\bibitem{equiv_form}
P.~Krokhmal, J.~Palmquist, and S.~Uryasev.
\newblock Portfolio optimization with conditional value-at-risk objective and constraints.
\newblock {\em Journal of Risk}, 4, 05 2003.

\bibitem{rockafellar2000optimization}
R.~T. Rockafellar, S.~Uryasev, et~al.
\newblock Optimization of conditional value-at-risk.
\newblock {\em Journal of risk}, 2:21--42, 2000.

\bibitem{schar2021decentralized}
F.~Sch{\"a}r.
\newblock Decentralized finance: On blockchain-and smart contract-based financial markets.
\newblock {\em FRB of St. Louis Review}, 2021.

\bibitem{vovk2013kernel}
V.~Vovk.
\newblock Kernel ridge regression.
\newblock In {\em Empirical Inference: Festschrift in Honor of Vladimir N. Vapnik}, pages 105--116. Springer, 2013.

\bibitem{wright2006numerical}
S.~J. Wright.
\newblock {\em Numerical optimization}.
\newblock 2006.

\end{thebibliography}

\end{document}